\documentclass[aps,showpacs,a4paper,floatfix,twocolumn,prc,amsmath,amssymb]{revtex4-1}
\usepackage{graphicx}
\usepackage{epsfig}
\usepackage{ulem}
\usepackage{morefloats}
\usepackage{rotating}
\usepackage{relsize}
\usepackage{color}

\begin{document}

\title{Neutron stars: from the inner crust to the core with the (Extended) Nambu-Jona-Lasinio model
}

\author{Helena Pais$^1$, D\'ebora P. Menezes$^2$, and Constan\c ca Provid\^encia$^1$}
\affiliation{$^1$CFisUC, Department of Physics, University of Coimbra, P-3004-516 Coimbra, Portugal \\
$^2$Departamento de F\'isica, Universidade Federal de Santa Catarina, Florian\'opolis, SC, CP. 476, CEP 88.040-900, Brazil}

\begin{abstract}
Nucleonic  matter is described within a su(2)  extended Nambu-Jona-Lasinio
  model.   Several parametrizations with different nuclear matter
  saturation properties are proposed.  At subsaturation, nuclear pasta
  phases are calculated within two methods: the coexistence-phases approximation and
the compressible liquid drop model, with the surface
tension coefficient determined using a geometrical approach  at zero temperature.
An unified equation of state of stellar matter for the inner crust, with the
  nuclear pasta phases,  and the  core is
  calculated.  The mass and radius of neutron stars within this
  framework are obtained for several families of hadronic and hybrid
stars.  The quark phase of hybrid stars is described within
the su(3) NJL model including a  vector term. Stellar macroscopic properties are in accordance with some of the recent results in the literature.
\end{abstract}
 

\maketitle

\section{Introduction} \label{I}

Neutron stars still remain today, despite several studies started many years ago, either from an observational point of view, or from a theoretical point of view, a big question mark, with respect to their constitution and characteristics. Currently, many efforts are being made to try to unravel a bit further our knowledge on this exotic objects, the remnants of core-collapse supernova events (see e.g. the forthcoming volume of EPJA Topical Issue on ``Exotic Matter in Neutron Stars'' \cite{Blaschke15epja}).

In the inner crust of neutron stars, at sub-saturation densities, a frustrated system, named nuclear pasta, appears due to the competition between the Coulomb and the strong forces. These exotic structures have their geometry changing as the density increases. One of the main interests on these nuclear shapes are the effect that they might have on the neutrino transport \cite{Horowitz-04I,Horowitz-04II,Sonoda-07,Alloy11}, and subsequent cooling of the neutron star. Recent studies \cite{Pons13} have shown evidence on the existence of the pasta phases, where these structures limit the maximum spin period of isolated X-ray pulsars.

The nuclear pasta phase has been studied under different assumptions, namely using semi-classical microscopic treatments, like quantum molecular dynamic calculations \cite{Watanabe-09,Horowitz-04II}, or using microscopic calculations, like 1D Hartree-Fock  \cite{Bonche-81, Bonche-82}, or 3D Hartree-Fock using SLy4 Skyrme model \cite{Magierski-02}.  Relativistic mean-field (RMF) calculations employing a model for the Lagrangian, which is based on microscopic Dirac-Brueckner-Hartree-Fock calculations, using realistic NN interactions \cite{Gogelein-08}, or using the Thomas-Fermi calculations, based on phenomenological energy-density functionals \cite{Avancini-08} are also widely used. RMF models do not consider explicit chiral invariance.
In this article, we use a different set of models for the description
of the  nucleonic homogeneous matter and the pasta phase: the
extended Nambu-Jona-Lasinio model with different parametrizations \cite{Koch87,
Bentz01,Moszkowski01,Burvenich03,ProvidenciaENJL03,Mishustin04,Moszkowski06},
where the chiral symmetry is included. 
An advantage of these models is the fact that, since they satisfy chiral symmetry, the EoS is also valid at higher densities, as the ones present in the center of compact objects. 

As discussed in \cite{Burvenich03}, a link between QCD and the description of nuclear matter and nuclei through effective hadronic fields can be established including QCD symmetries in the Lagrangian density of the system. Using the extended NJL (eNJL) model developed in the present study, we include chiral symmetry, together with the mechanism of mass generation and  binding of nuclear matter, the chiral condensates being built from nucleonic degrees of freedom. 

The pasta phase calculation (see e.g. \cite{Pais-15} and references
therein) is done by  considering two different methods: the
coexisting-phases (CP) approximation, where the Gibbs equilibrium
conditions are used to get the lowest free-energy state, and the
surface and Coulomb terms are added ``by hand'', and the compressible
liquid drop (CLD) model, where, unlike the CP approximation, both the
Coulomb and surface terms  are taken into account in the minimization of the total energy of the system.

In the present paper, our aim is the complete description of possible
matters in the interior of neutron stars.  Once matter at
subsaturation and suprasaturation densities are  obtained and
understood within the su(2) eNJL,  we consider also the  su(3) version
of the NJL model \cite{NJLa,NJLb,Vogl91,Klevansky92,Buballa05} with a
vector interaction \cite{Menezes14,Hanauske01,Pagliara08,Bonanno12}  
to  describe a possible neutron star core in a hybrid star.
We construct the complete stellar EoS by considering the BPS EoS \cite{bps} for
the outer crust, the pasta EoS for the inner crust, and investigate
two different scenarios at high  densities: i)
nucleonic matter and ii) nucleonic and quark matter via Maxwell construction. In the second
case, we investigate the possibility of a hybrid star with a quark
core. Whenever stellar matter is considered, either in the pasta phase
or in its center, $\beta-$equilibrium and charge neutrality are
enforced.  The procedure just described allows us to obtain the
  inner crust and the core stellar matter  EoS, within the same framework.

The paper is organized as follows: the formalism is briefly reviewed
in Sec. \ref{II} that is divided in different subsections. We first
analyze homogeneous nucleonic matter and its saturation properties
obtained with the su(2) eNJL model and then briefly outline the main
aspects of the construction of the pasta phase, including the surface
tension calculation. The su(3) NJLv model is then introduced for
  the description of quark matter, and for the calculation of the hybrid star mass and radius.
Section \ref{III} is devoted to the presentation and discussion of the results, while the main conclusions are given in Sec. \ref{IV}.

\section{Formalism} \label{II}

In this section, we present the model that describes the nucleonic
equation of state (EoS), and make a review of the different formalisms
needed to describe the sub-saturation and the quark EoS.

\subsection{eNJL model}

 The nucleonic NJL model can be extended 
 to yield reasonable saturation properties of nuclear matter, the field
$\psi$ being the nucleon field \cite{Koch87,
Bentz01,Moszkowski01,ProvidenciaENJL03,Mishustin04,Moszkowski06}. 
An effective density
dependent coupling constant is obtained if the following extended
NJL (eNJL) Lagrangian density, which actually pushes chiral symmetry
restoration to higher densities, is considered:
\begin{eqnarray}
{\cal L}&=&\bar\psi(i\gamma^\mu\partial_\mu-m)\psi+G_s[(\bar\psi\psi)^2
+(\bar\psi i\gamma_5\vec\tau\psi)^2] \nonumber\\
&-&G_v(\bar\psi\gamma^\mu\psi)^2-G_{sv}[(\bar\psi\psi)^2+(\bar\psi i\gamma_5\vec\tau\psi)^2]
(\bar\psi\gamma^\mu\psi)^2 \nonumber \\
&-&G_\rho \left[(\bar\psi\gamma^\mu \vec\tau\psi)^2+ (\bar\psi \gamma_5\gamma^\mu\vec\tau\psi)^2\right]  \\
&-& G_{v\rho}(\bar\psi\gamma^\mu\psi)^2\left[(\bar\psi\gamma^\mu \vec\tau\psi)^2+ (\bar\psi \gamma_5\gamma^\mu\vec\tau\psi)^2\right] \nonumber \\
&-& G_{s\rho}\left[(\bar\psi\psi)^2+(\bar\psi i\gamma_5\vec\tau\psi)^2\right]\left[(\bar\psi\gamma^\mu \vec\tau\psi)^2+ (\bar\psi \gamma_5\gamma^\mu\vec\tau\psi)^2\right]. \nonumber
\label{lagran}
\end{eqnarray}
For nuclear matter,  the degeneracy is
$\nu=2N_f$, and  $\Lambda$ is such that $M=939$ MeV is the nucleon mass
in the vacuum, determined variationally.
The term in $G_v$ simulates a chiral
invariant short range repulsion between nucleons. 
The term in $G_{sv}$ accounts for the density dependence of the
scalar coupling. 
For nuclear matter, the NJL model leads to binding, but
the binding energy per particle does not have a minimum except
at a rather high density where the nucleon mass is small or vanishing. 
The introduction of  the  $G_{sv}$ coupling term
is required to correct this. The isovector-vector term
(the $G_\rho$ term)  allows the  description of isospin asymmetric nuclear
matter.  A current mass $m$ term that breaks explicitly the
  chiral symmetry is introduced in some parametrizations to
  make the restoration of the chiral symmetry less abrupt.  The
  terms $G_{\omega\rho}$ and $G_{s\rho}$ make the symmetry energy softer.

\begin{table*}[!htbp]
  \centering
  \caption{The coupling constants of the models discussed in the
    present work.} 
  \begin{tabular}{c c c c c c c c c}
    \hline
    \hline
	Model & $G_s$ (fm$^2$) & $G_v$ (fm$^2$) & $G_{sv}$ (fm$^8$) & $G_\rho$ (fm$^2$) & $G_{v\rho}$ (fm$^8$) & $G_{s\rho}$ (fm$^8$) & $\Lambda$ (MeV) & $m$ (MeV)  \\
    \hline
	eNJL1 &	4.855 &	4.65 & -6.583 & 0.5876 & 0 & 0& 388.189 & 0 \\
	eNJL1$\omega\rho$1 &	4.855 &	4.65 & -6.583 & 0.5976 & -1 & 0 & 388.189 & 0 \\
	eNJL1$\omega\rho$2 &	4.855 &	4.65 & -6.583 & 0.6476 & -6 & 0 & 388.189 & 0\\
	eNJL2 &	3.8 &	3.8 & -4.228 & 0.6313 & 0 & 0& 422.384 & 0  \\
	eNJL2$\omega\rho$1 &	3.8 &	3.8 & -4.228 & 0.6413 &-1 & 0 & 422.384 &0 \\
	\hline
	eNJL3 &	1.93 &	3. & -1.8 & 0.65 & 0 & 0 & 534.815 & 0 \\
	eNJL3$\sigma\rho$1 &	1.93 &	3. & -1.8 & 0.0269 & 0 & 0.5 & 534.815 & 0 \\
	\hline
	eNJL1m&	1.3833 &	1.781 & -2.943 & 0.7 & 0 & 0 & 478.248 & 450 \\
	eNJL1m$\sigma\rho$1 & 1.3833 & 1.781 & -2.943 & 0.0739 & 0 & 1 & 478.248 & 450 \\
	eNJL2m&	1.078 &	1.955 & -2.74 & 0.75 & 0 & 0 & 502.466 & 500 \\
	eNJL2m$\sigma\rho$1 &	1.078 &	1.955 & -2.74 & -0.1114 & 0 & 1 & 502.466 & 500 \\
	\hline   
    \hline
  \end{tabular}
 \label{tab1}
\end{table*}

The  thermodynamical potential per volume corresponding to (\ref{lagran}) is given by
\begin{eqnarray}
\omega(\mu)&=&\varepsilon_{kin} + m\rho_s
-G_s\rho_s^2 + G_v\rho^2+G_{sv}\rho_s^2\rho^2 +G_\rho\rho_3^2 \nonumber \\
&+&G_{v\rho}\rho^2\rho_3^2 +G_{s\rho}\rho_s^2\rho_3^2
-\mu_p\rho_p-\mu_n\rho_n \, ,
\label{omega}
\end{eqnarray}
where exchange terms have been neglected. The kinetic energy density
is defined as 
\begin{eqnarray}
\varepsilon_{kin}&=&\langle\bar\psi(\vec\gamma\cdot\vec p)\psi\rangle = F_1(M,k_{F_i})-F_1(M,\Lambda)\, , \nonumber \\
F_1(M,x)&=&\int_0^x \frac{dp}{\pi^2}p^2\sqrt{M^2+p^2} \, ,
i=p,n ,
\end{eqnarray}
and $\rho, \rho_s$ and $\rho_3$ are the baryonic, scalar and isovector
densities, respectively, and are given by $\rho=\rho_p+\rho_n$,
$\rho_s=\rho_{sp}+\rho_{sn}$ and $\rho_3=\rho_p-\rho_n$.
The proton and neutron densities and scalar densities are given by the usual expressions
\begin{equation}
\rho_i=\int_0^{k_{F_i}} \frac{dp}{\pi^2} p^2  \, 
\end{equation}
and
\begin{eqnarray}
\rho_{s_i}&=&M\left[F_0(M,k_{F_i})-F_0(M,\Lambda)\right] \, , \nonumber \\
F_0(M,x)&=&\int_0^x \frac{dp}{\pi^2}\frac{p^2}{\sqrt{M^2+p^2}} \, ,
i=p,n .
\end{eqnarray}

The pressure of the system is given by $P=-\omega(\mu)+\varepsilon_0$, and the total energy density is given by $\varepsilon=-P+\mu_p\rho_p+\mu_n\rho_n$, with $\varepsilon_0$ being the energy density in the vacuum.
The condition
$\partial\omega/\partial M=0$ determines the effective nucleon mass
given by:
\begin{equation}
M=m-2G_s\rho_s+2G_{sv}\rho_s\rho^2+2G_{s\rho}\rho_s\rho_3^2.
\label{mass}
\end{equation}

The free nucleon mass, $M_0$, is the value of $M$ at zero
  chemical potential. 
The conditions $\partial\omega/\partial p_{F_i}=0$ determine the
chemical potentials
\begin{eqnarray}
\mu_p&=&E_{p_{F}}^p+2G_v\rho+2G_{sv}\rho\rho_s^2+2G_\rho\rho_3+2G_{v\rho}\rho_3^2\rho \nonumber \\
&+&2G_{v\rho}\rho^2\rho_3+2G_{s\rho}\rho_3\rho_s^2, \\
\mu_n&=&E_{p_{F}}^n+2G_v\rho+2G_{sv}\rho\rho_s^2-2G_\rho\rho_3+2G_{v\rho}\rho_3^2\rho \nonumber \\
&-&2G_{v\rho}\rho^2\rho_3-2G_{s\rho}\rho_3\rho_s^2,
\end{eqnarray} 
with $E_{p_F}^i=\sqrt{M^2+p_F^{i2}}$, $i=p,n$.
These conditions together with Eq. (\ref{mass}) fix the values of $p_F^i,\,M$ for given $\mu_i.$ 

For reference and to help the discussion,  we show in Table \ref{tab1}
the coupling constants, and in Table \ref{tab2} the symmetric nuclear
matter properties for the models we are using in this study, eNJLx,
eNJLx$\omega\rho$y and eNJLJx$\sigma\rho$y  (without current mass), and eNJLxm and eNJLxm$\sigma\rho$y (with current mass). Models eNJLx$\omega\rho$y (eNJLx$\sigma\rho$) contain the $\omega\rho$ ($\sigma\rho$) coupling term in the Lagrangian density, i.e., $G_{v\rho} \neq 0$ ($G_{s\rho} \neq 0$). 
We have fixed the symmetry energy at $\rho=0.1$ fm$^{-3}$ at the same value obtained for eNJLx (eNJLxm), and we calculated the new $G_\rho$ constants, by fixing the $G_{v\rho}$ ($G_{s\rho}$) coupling constant.

\begin{table}[!htbp]
  \centering
  \caption{Symmetric nuclear matter properties at saturation density $\rho_0$ (energy per particle $B/A$, incompressibility $K$, symmetry energy $E_{sym}$ and symmetry energy slope $L$). All the quantities are in MeV, except for $\rho_0$, given in fm$^{\rm {-3}}$.} 
  \begin{tabular}{c c c c c c c c c c c}
    \hline
    \hline
	Model & \phantom{a} & $\rho_0$ & \phantom{a} & $B/A$ & \phantom{a} & $K$ &\phantom{a} &$E_{sym}$ & \phantom{a}&$L$   \\
    \hline
	eNJL1&\phantom{a} &	0.148 	& \phantom{a}&-16.34 &\phantom{a}	& 267.26 & \phantom{a}&33.0 & \phantom{a}&99.90 \\
	eNJL1$\omega\rho$1 &	\phantom{a}&0.148 &	\phantom{a}&-16.34 &	\phantom{a}& 267.26 & \phantom{a}&32.65& \phantom{a}&95.02 \\	
	eNJL1$\omega\rho$2 &	\phantom{a}&0.148 &	\phantom{a}&-16.34 &	\phantom{a}& 267.26 & \phantom{a}&30.91& \phantom{a}&70.61 \\	
	eNJL2&\phantom{a} &	0.148 	& \phantom{a}&-15.56 &\phantom{a}	& 231.13 & \phantom{a}&33.0 & \phantom{a}&95.03 \\
	eNJL2$\omega\rho$1 &	\phantom{a}&0.148 &	\phantom{a}&-15.56 &	\phantom{a}& 231.13 & \phantom{a}&32.65 & \phantom{a}&90.15 \\
	\hline
	eNJL3&\phantom{a} &	0.148 	& \phantom{a}&-15.69 &\phantom{a}	& 239.70 & \phantom{a}&31.65 & \phantom{a}&85.26 \\
	eNJL3$\sigma\rho$1&\phantom{a} &	0.148 	& \phantom{a}&-15.69 &\phantom{a}	& 239.70 & \phantom{a}&29.91 & \phantom{a}&64.45\\
	\hline
	eNJL1m&\phantom{a} &	0.148 	& \phantom{a}&-16.05 &\phantom{a}	& 233.75 & \phantom{a}&32.46 & \phantom{a}&86.20 \\
	eNJL1m$\sigma\rho$1&\phantom{a} &	0.148 	& \phantom{a}&-16.05 &\phantom{a}	& 233.75 & \phantom{a}&30.28 & \phantom{a}&60.32 \\
	eNJL2m&\phantom{a} &	0.148 	& \phantom{a}&-16.22 &\phantom{a}	& 286.63 & \phantom{a}&33.66 & \phantom{a}&89.20 \\
	eNJL2m$\sigma\rho$1&\phantom{a} &	0.148 	& \phantom{a}&-16.22 &\phantom{a}	& 286.63 & \phantom{a}&31.13 & \phantom{a}&59.04\\
	\hline   
    \hline
  \end{tabular}
 \label{tab2}
\end{table}

\subsection{The coexisting-phases approximation and the compressible liquid drop model} \label{CP-CLD}

In order to describe the nonuniform $npe$ matter inside the Wigner-Seitz unit cell,
which is taken to be a sphere (bubble), a cylinder (tube), or a slab, in three, two,
and one dimensions, we use two different methods: the coexistence-phases (CP) approximation and the compressible liquid drop (CLD) model.  In the CP approximation, matter is organized into separated regions of higher and lower density; the higher
ones being the pasta phases, and the lower ones being a
background nucleon gas. The interface between these regions
is sharp and finite-size effects are taken into account by surface
and Coulomb terms in the energy density \cite{Avancini-12Cl}. The Gibbs equilibrium conditions are imposed to get the lowest-energy state and, for a temperature $T=T^I=T^{II}$ and a fixed proton fraction, are given by 
\begin{eqnarray}
\mu_n^I=\mu_n^{II}, \nonumber \\
\mu_p^I=\mu_p^{II}, \nonumber \\
P^I=P^{II}, \nonumber
\end{eqnarray}
where $I$ and $II$ label the high- and low-density phases, respectively. After the lowest energy state is achieved, the surface and Coulomb terms are added to the total energy density of the system, which is given by

\begin{eqnarray}
\varepsilon=f\varepsilon^I+(1-f)\varepsilon^{II}+\varepsilon_e+\varepsilon_{surf}+\varepsilon_{Coul},
\end{eqnarray}
where $f$ is the volume fraction of phase $I$. 

In the CLD model \cite{Lattimer-91, Baym-71, Lattimer-85, bao14}, the equilibrium conditions of the system are derived from the minimization of the total free energy \cite{Baym-71}, including the surface
and Coulomb terms. The equilibrium conditions for a fixed proton
  fraction become
\begin{eqnarray}
\mu_n^I&=&\mu_n^{II}, \nonumber \\
\mu_p^I&=&\mu_p^{II}-\frac{\varepsilon_{surf}}{f(1-f)(\rho_p^I-\rho_p^{II})}, \nonumber \\
P^I&=&P^{II}-\varepsilon_{surf}\Big(\frac{1}{2\alpha}+\frac{1}{2\Phi}\frac{\partial\Phi}{\partial f}-\frac{\rho_p^{II}}{f(1-f)(\rho_p^I-\rho_p^{II})}\Big), \nonumber 
\end{eqnarray}
where $\alpha=f$ for droplets, rods and slabs, and $\alpha=1-f$ for tubes and bubbles. $\Phi$ is given by
\begin{eqnarray}
\Phi=\left\lbrace\begin{array}{c}
\left(\frac{2-D \alpha^{1-2/D}}{D-2}+\alpha\right)\frac{1}{D+2}, D=1,3 \\
\frac{\alpha-1-\ln \alpha}{D+2}, D=2 \quad .
\end{array} \right. 
\end{eqnarray}
For more details on both methods, the reader should refer to \cite{Pais-15} and references therein.

\subsection{The surface tension}
\begin{figure}[!htbp]
   \includegraphics[width=0.45\textwidth]{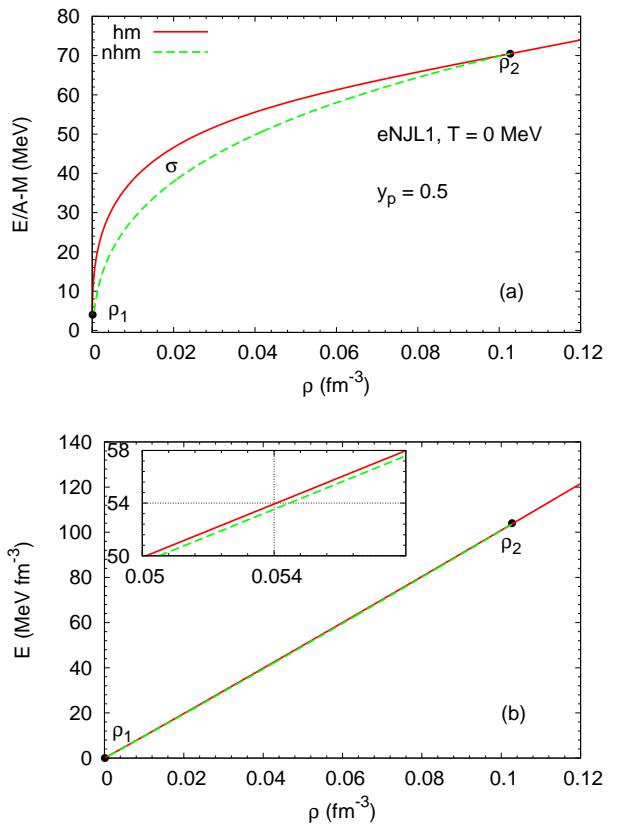}\\
	\caption{(Color online) Energy per baryon (a) and energy density (b) versus density, where the two coexistence points are shown, for homogeneous (hm) and non-homogeneous (nhm) matter.} 
\label{fig1}
\end{figure}

We use a geometrical approach to obtain a numerical value for the
surface tension coefficient. This method was introduced and discussed
in \cite{Pinto12} for quark matter.
The surface tension coefficient, $\sigma$, is given by
\begin{eqnarray}
\sigma=\frac{a}{\rho_g}(2\varepsilon_g)^{1/2}\int_{\rho_1}^{\rho_2}(\Delta\varepsilon)^{1/2}d\rho,
\end{eqnarray}
with $\rho_g=\frac{\rho_1+\rho_2}{2}$, $\varepsilon_g=\frac{\varepsilon(\rho_1)+\varepsilon(\rho_2)}{2}$, and $\Delta\varepsilon$ the difference between the energy density of homogeneous matter and the non-uniform matter, given by  $\Delta\varepsilon=\varepsilon_{hm}-\varepsilon_{nhm}$. These energy densities were fitted to a functional form given by $\varepsilon=b_0+b_1\rho+b_2\rho^2$. $\rho_1$ and $\rho_2$ are the two coexistence points. In Figure \ref{fig1}, we show the energy per baryon and the energy density as a function of the density for the homogeneous and non-homogeneous cases. The surface tension, $\sigma$,  which measures  the energy per unit area necessary to create a planar interface between the two phases,  is defined in terms of the EoS, as in \cite{Pinto12}. The width of the interface region and magnitude of $\sigma$ are controlled by the adjustable parameter $a$. Here $a$ was chosen to be 0.1 so that it reproduces the surface tension coefficient for the NL3 model \cite{Lalazissis-97} within a Thomas-Fermi calculation \cite{Avancini-10Cl}, for a fixed proton
fraction of $y_p=0.5$. We also tested for a different RMF model, the
TW model \cite{Typel-99}, and we obtained a similar result, $a=0.13$, for a fixed
proton fraction of 0.5. Since this parameter $a$ depends on the isospin, we
calculated it for several values of the proton fraction, and then we
fitted it to a functional $a=a_1 +a_2x^2+a_3x^4+a_4x^6$, in order to
calculate the pasta phase in $\beta-$equilibrium matter. We have
obtained $a_1=-0.00391407, a_2=0.251366, a_3=5.5648, a_4=-18.5799$, taking NL3 as reference.

\subsection{NJLv model} 
 
 \begin{figure*}[!htbp]
   \includegraphics[width=0.8\textwidth]{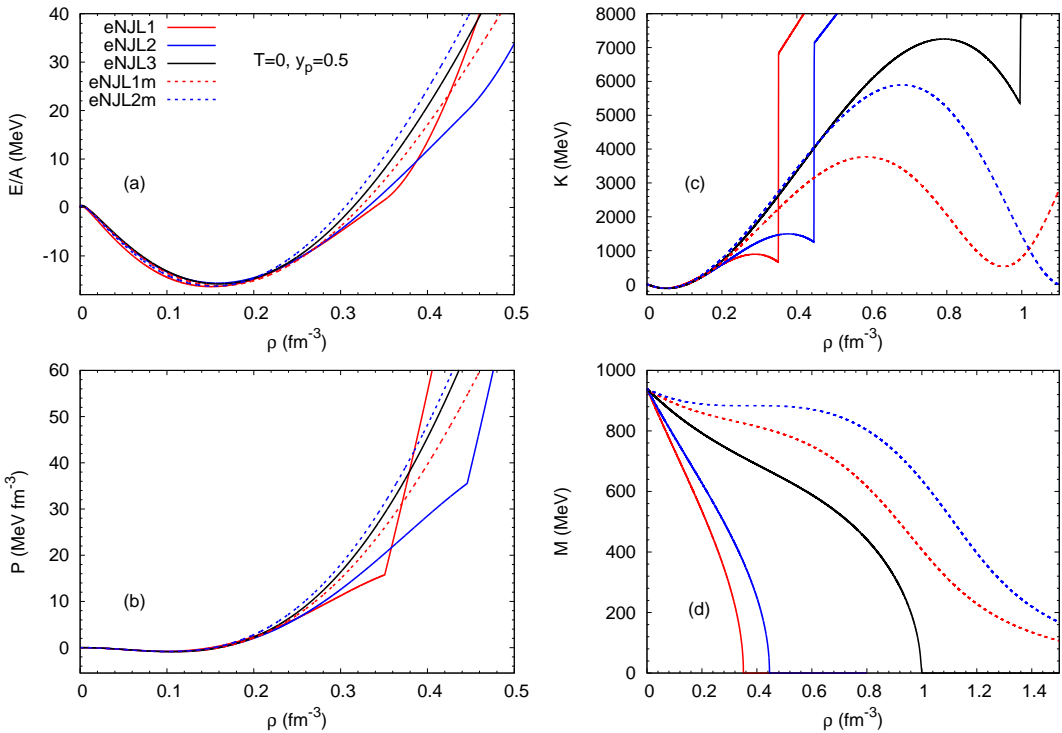}
   \caption{(Color online) Symmetric nuclear matter properties  as a function of the density for all the models considered in this study. }
\label{fig2}
\end{figure*}

 The su(3) NJL model \cite{NJLa,NJLb,Vogl91,Klevansky92,Buballa05,Menezes14} is given by the following Lagrangian density 
  \begin{eqnarray}
 {\cal L}&=&{\cal L}_f+{\cal L}_l+{\cal L}_{sym}+{\cal L}_{vec}+{\cal L}_{det}
 \end{eqnarray}
 with
 \begin{eqnarray}
 {\cal L}_f&=&\bar\psi_f(i\gamma_\mu\partial^\mu-m_f)\psi_f ,\quad 
{\cal L}_l=\bar\psi_l(i\gamma_\mu\partial^\mu-m_l)\psi_l \, , \nonumber \\
 {\cal L}_{sym}&=&G_s\sum_{a=0}^8[(\bar\psi_f\lambda_a\psi_f)^2+(\bar\psi_f i\gamma_5\lambda_a\psi_f)^2]\, , \nonumber \\ 
{\cal L}_{vec}&=&-G_v\sum_{a=0}^8[(\bar\psi_f\gamma^\mu\lambda_a\psi_f)^2+(\bar\psi_f\gamma^\mu\gamma_5\lambda_a\psi_f)^2] \, , \nonumber \\ 
{\cal L}_{det}&=&-G_t\lbrace det_f[\bar\psi_f(1+\gamma_5)\psi_f]+det_f[\bar\psi_f(1-\gamma_5)\psi_f]\rbrace  \, . \nonumber
 \end{eqnarray}
 
 Here $\psi_f$ is the 3-flavor quark field. The effective mass, $M_i$, is given by
$$M_i=m_i-4G_s\rho_{s_i}-2G_t\rho_{s_j}\rho_{s_k} \, ,
$$
with $(i,j,k)$ being any permutation of $(u,d,s)$, and the chemical potentials by
$$\mu_i=E_{i_{F}}+4G_v\rho_i$$ 
with $E_{i_F}=\sqrt{M_i^2+p_F^{i2}}$, $i=u,d,s$.
These conditions fix the values of $p_F^i,\,M$ for given $\mu_i.$ 

When applying the su(3) NJLv model to describe the core of hybrid
  stars in Section \ref{ns}, we use a new parametrization, recently
  proposed in  \cite{Pereira16}, which was built by considering the
  quark mass in the vacuum equal to 313 MeV, and meson properties in
  the vacuum. 
This allows both the hadronic and quark models to have the same
nucleon mass, while constructing hybrid stars. 
We consider four different parameter sets, the difference being the
strenght of the coupling of the vector interaction. The scalar coupling, $G_s=1.781/\Lambda^2$ MeV$^{-2}$, the t'Hooft interaction constant, $G_t=-9.29/\Lambda^5$ MeV$^{-5}$, and the cut-off parameter, $\Lambda=630$ MeV, are equal for all the four parametrizations. The coupling for the vector interaction, $G_v$, is given as $G_v=xG_s, x=0,0.05,0.12,0.2$. We will be calling our set of models NJL${i}, i=1,2,3,4$, for $x=0,0.05,0.12,0.2$, respectively.

\subsubsection{The Maxwell construction}

We consider different EoS for the construction of a hybrid
star: a hadronic EoS, where we use  the parameter sets presented in Tables  \ref{tab1} and \ref{tab2}, and a quark EoS, where we consider the NJL$i$ models presented in the first section, and we perform a Maxwell construction. This construction says that two phases are in equilibrium when their chemical potentials, temperatures and pressures are equal:
\begin{eqnarray}
T_H=T_Q=0 \\
\mu_H=\mu_Q \\
P_H(\mu)=P_Q(\mu)
\end{eqnarray}
Finding the pressure at the transition, $P_t$, will then give us a
range of densities with  a mixed phase. 

With the complete EoS, we  integrate the TOV
\cite{tolman39,oppenheimer39} equations and find the mass-radius
relation for the family of stars. This is a simplified approach to
the description of the deconfinement phase transition, but it has been
shown that if the surface tension of a quark droplet immersed in
nuclear matter is high enough, the results obtained are quite realistic
\cite{Yasutake14}.

\section{Results} \label{III}

In the present section we use the model presented in the previous
sections to describe stellar dense  matter as found in neutron
stars. In particular, we present a unified EoS, except for the
outer crust, which is, however, essentially constrained by observational data.
We fix the parameters of the model constrained by the properties
of nuclear matter at saturation. Properties of symmetric matter and
pure neutron matter are discussed. It has been shown that
in order to get consistently the radius of stars with masses $M \lesssim 
1.5 M_\odot$, it is important to describe the inner crust in an
appropriate way. We, therefore, calculate the inner crust EoS and
discuss how the properties of the EoS affect the  structure of the
pasta phases. Finally, we build the EoS of $\beta$-equilibrium
matter and integrate the TOV equations in order to get the mass
versus radius curves. We consider both nucleonic stars and hybrid
stars. The EoS for hybrid stars takes into account a possible
deconfinement phase transition to quark matter. Quark matter is 
described within the su(3) NJL$i$, discussed above, taking into account the possibility of
the strangeness onset. 

\begin{figure*}[!htbp]
   \includegraphics[width=0.8\textwidth]{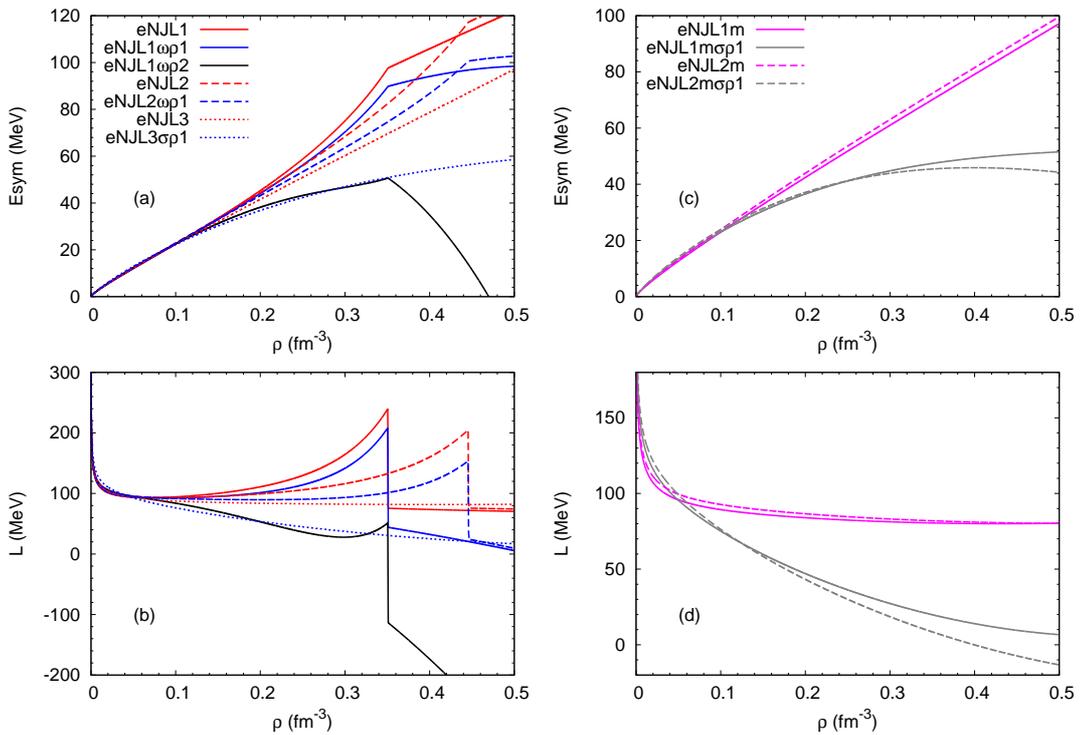}
   \caption{(Color online) Symmetry energy, $E_{sym}$, (a) and (c), and its slope, $L$, (b) and (d), for the models considered in this study. } 
\label{fig3}
\end{figure*}

\subsection{Equation of state for symmetric matter}

We consider the parametrizations eNJL1, eNJL2, eNJL3 and eNJL1m and
eNJL2m, as referred before, with
 their parametrizations given in Table \ref{tab1} and their properties
 for symmetric nuclear matter 
in Table \ref{tab2}.
For eNJL1, eNJL2, and eNJL3, the symmetry energy properties at saturation are very similar, but their
 isoscalar properties differ slightly, eNJL2 having a smaller
 incompressibility. Their values are, however, all within the interval
 proposed in \cite{Khan13}, $K=230\pm 40$ MeV. Another property that
 distinguishes these three models is the density of chiral symmetry
 restoration, respectively, 0.35, 0.45, and 0.998 fm$^{-3}$, for eNJL1, eNJL2, and
 eNJL3, see Fig. \ref{fig2}, bottom right panel. As we see next, this
 allows the construction of  EoS  for hadronic matter with different behaviors at intermediate and high densities.

\begin{figure*}[!htbp]
	\begin{tabular}{cc}
	   \includegraphics[width=0.5\textwidth]{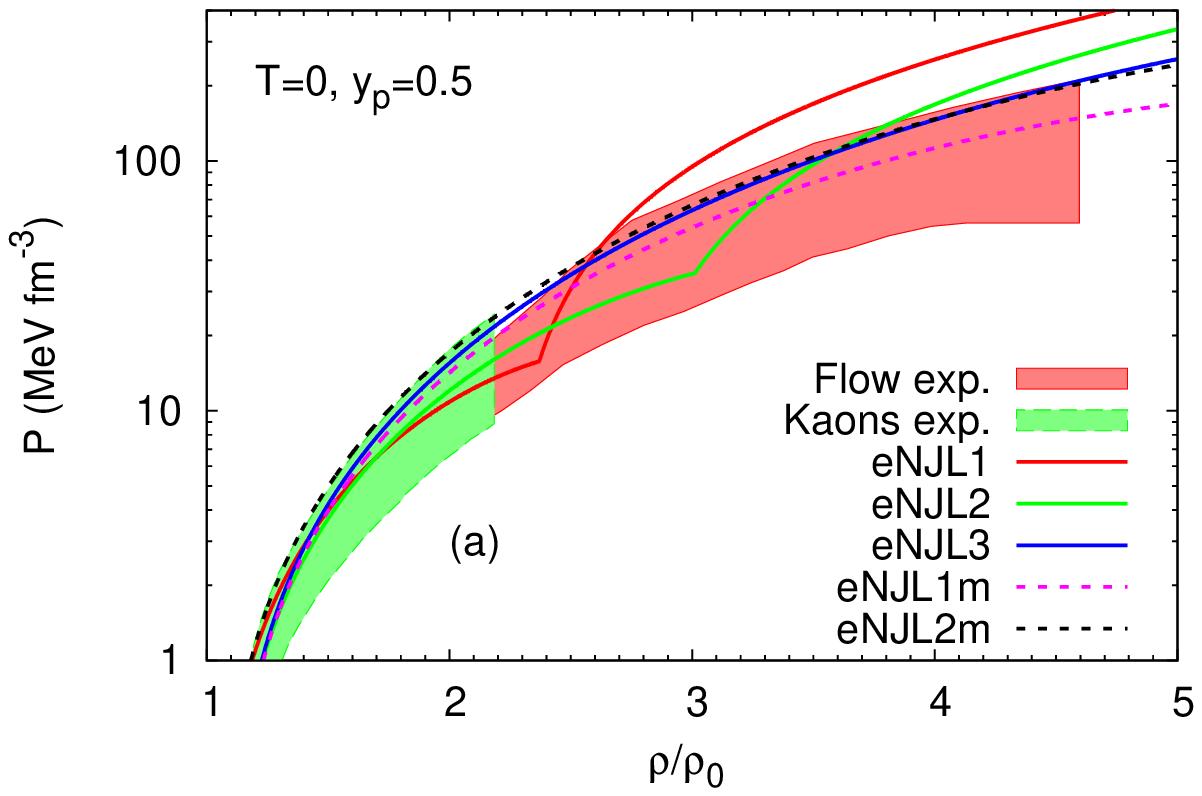}
    		\includegraphics[width=0.5\textwidth]{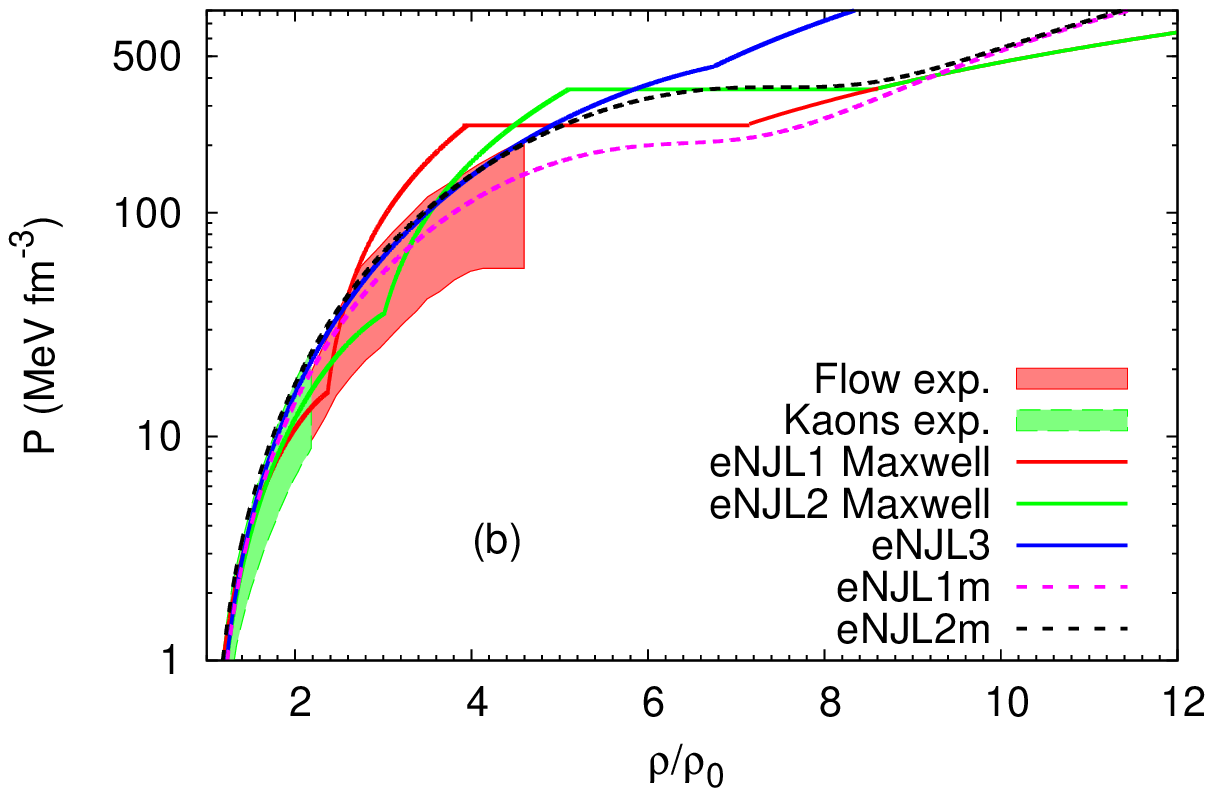}\
    	\end{tabular}	
\caption{(Color online) Pressure  as a function of the density for symmetric nuclear matter and for all the models considered in this study. The red region represents the experimental results from Danielewicz et al and the green region from the Kaons experiment. In the right panel, (b), a Maxwell construction for the eNJL1 and eNJL2 models is also shown.}    
\label{fig4}
\end{figure*}

In Fig. \ref{fig2}, the energy per particle, and pressure (left panels), and
the incompressibility and effective mass (right panels) of symmetric nuclear matter are plotted as a
function of density. The main feature of these EoS is the change that
occurs at the chiral symmetry restoration density, $\rho_\chi$:  the EoS
becomes much stiffer after the transition, which, for eNJL3, only happens at $\rho\sim 7\rho_0$. Just before the transition,
there is a softening clearly seen in the incompressibility, $K$,
followed by a change of the slope of the pressure, and a discontinuity
of the  incompressibility. The models with current mass do not present this feature.  With a proper choice of
  the parameters, the chiral symmetry restoration takes place at
  higher densities, and the EoS becomes much softer at intermediate
  densities. As we will see later, the stiffening of the
EoS will allow for very massive stars.

 eNJLx and eNJLxm have a quite large slope of the symmetry energy at
saturation density, larger than laboratory constraints seem to
impose \cite{Steiner16}. We, therefore, built six other
parametrizations with a smaller slope, $L$, by including in the
Lagrangian density a mixed  vector-isoscalar - vector-isovector (scalar-isoscalar - vector-isovector)
eight point  term. These parametrizations are designated by eNJLx$\omega\rho$y  (eNJLx(m)$\sigma\rho$y).
The symmetry energy and its slope at saturation density of all these
models are shown as a function of the density in
Fig.\ref{fig3}. While below $\rho_\chi$, the  eNJL1, eNJL2, and eNJL3
models behave as relativistic mean field (RMF) models, see
\cite{Ducoin-08, Ducoin11}, above this density, the symmetry  energy
becomes much softer and its slope may even become immediately
negative. This occurs for all models with the mix vector-isoscalar - vector-isovector term.
The symmetry energy will eventually become negative at some density
above $\rho_\chi$, indicating an instability and a tendency for
stellar matter to become  pure neutron matter.  If the restoration
  of chiral symmetry occurs at a density not high enough, the model
  becomes inadequate to describe stellar matter. In the last subsection,  only models which do not
  predict a neutron instability will be used to describe stellar matter.

The models with current mass (right panels) do not show 
such an abrupt behavior as they have a softer restoration of the chiral symmetry. For these models, we also consider a term that  couples the chiral condensate  and the isospin density, to make the symmetry
 energy less hard, allowing   the parametrizations to satisfy
 constraints from microscopic calculations based on chiral NN and 3N
 interactions  \cite{Hebeler-13} and Quantum Monte Carlo  results
 \cite{Gandolfi12}. At the same time, we avoid that the symmetry
 energy becomes negative below a reasonably high density.  This
   behavior is possible by including a coupling of the isovector-vector
   term to a isoscalar-scalar term, and not to a isoscalar-vector
   term, because at high densities the chiral condensate weakens.

 In Fig. \ref{fig4}, we compare the symmetric nuclear matter
   pressure to experimental results obtained from collective flow data in
heavy-ion collisions \cite{Danielewicz02} and from the KaoS experiment
\cite{Lynch09,Fuchs06,Sagert12,Fuchs01}. The left panel only contains
nucleonic EoS.  All EoS shown in this panel satisfy the constraints imposed by the KaoS
experiment, which refer to densities from above the saturation density to twice the
saturation density. The  models eNJL2 and e NJL3 are also compatible with
constraints from HIC flow experiments. The eNJL1 model, however, 
becomes quite hard above 2.5 $\rho_0$, even considering that
the constraints imposed in \cite{Danielewicz02} are too stringent,
since the data analysis contains some model dependence.

We have allowed for a possible deconfinement phase transition in
models eNJL1 and eNJL2. The symmetric matter EoS obtained within a
Maxwell construction are plotted in the right panel of
Fig. \ref{fig4}. We observe that, for both models, the
deconfinement phase transition occurs above the range of densities
constrained by the flow data, and, therefore, the conclusions are
similar to the ones derived from the left panel. However, for
densities above 4$\rho_0$, deconfinement gives rise to a clear
softening of the EoS.  For comparison, we replot in the right panel the  EoS of
models eNJL1m, eNJL2m and eNJL3.

\begin{figure}[!htbp]
   \includegraphics[width=0.45\textwidth]{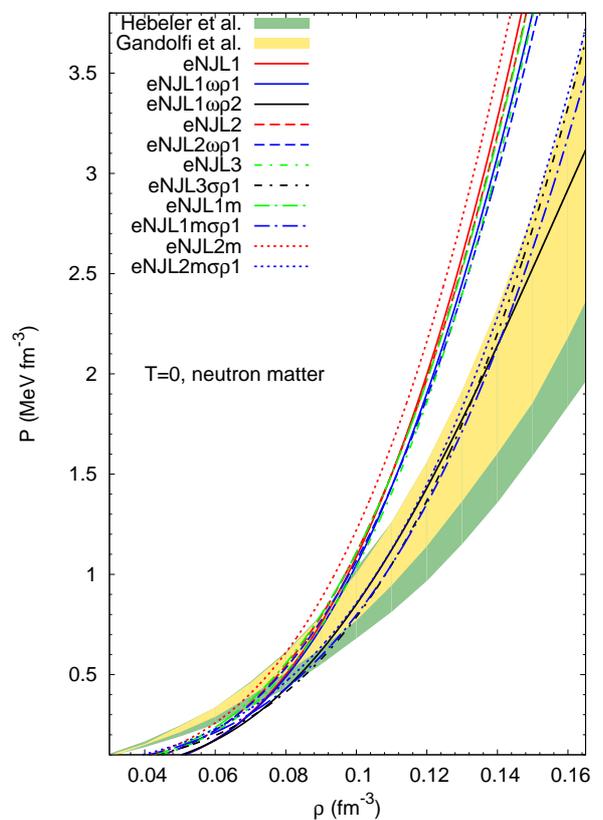}
   \caption{(Color online)  Pressure as a function of the density for
     neutron matter for the eNJLx, eNJLx$\omega\rho$y and
     eNJLx(m)$\sigma\rho$y interactions considered in this study. The
     colored bands are the results from \cite{Hebeler-13} (green) and
     \cite{Gandolfi12} (gold).}    
\label{fig5}
\end{figure}

\begin{figure*}[!htbp]
\begin{tabular}{cc}
   \includegraphics[width=0.45\textwidth]{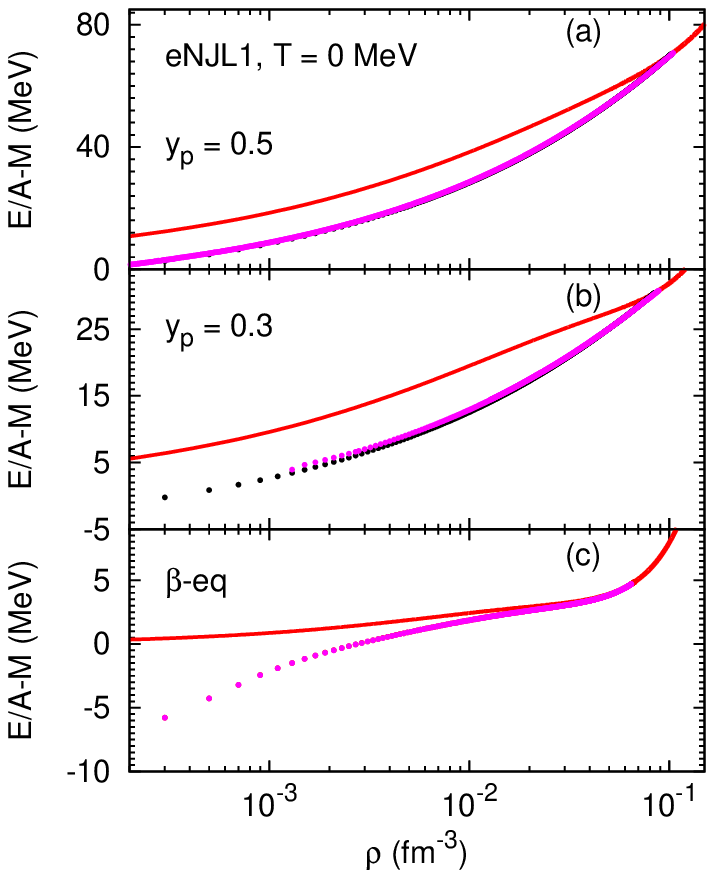}
   \includegraphics[width=0.45\textwidth]{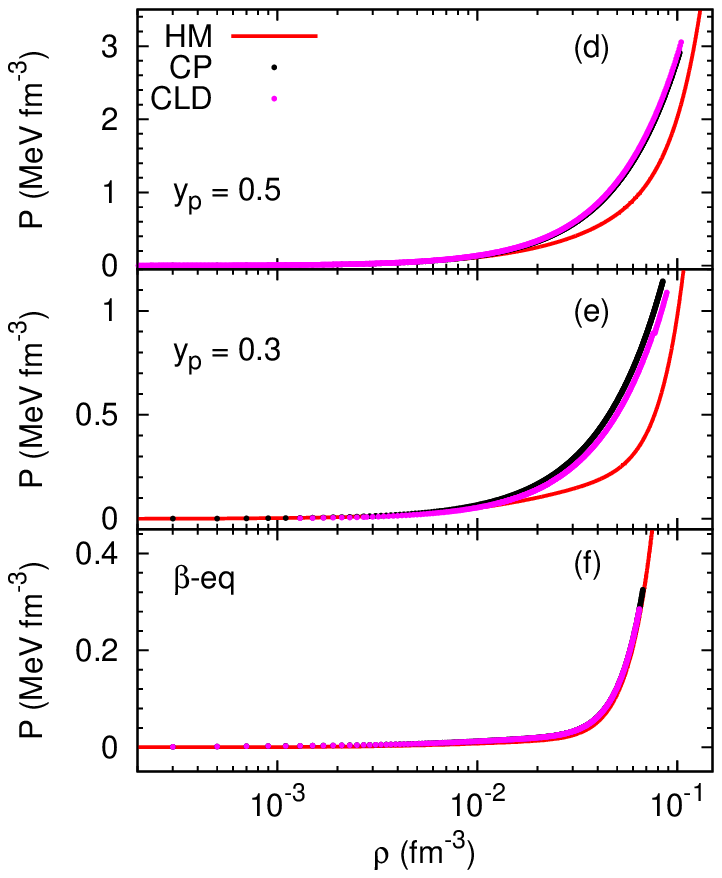}
\end{tabular}
   \caption{(Color online) Energy per baryon, and  pressure, as a function of the density for the eNJL1 interaction, for $\beta-$equilibrium matter, (c) and (f), $y_p=0.3$, (b) and (e), and $y_p=0.5$, (a) and (d), proton fractions. Results with pasta (within the CP (black dots) and CLD (pink dots) approaches) and for homogeneous matter (red solid line) are shown.}    
\label{fig6}
\end{figure*}

\begin{figure}
   \includegraphics[width=0.45\textwidth]{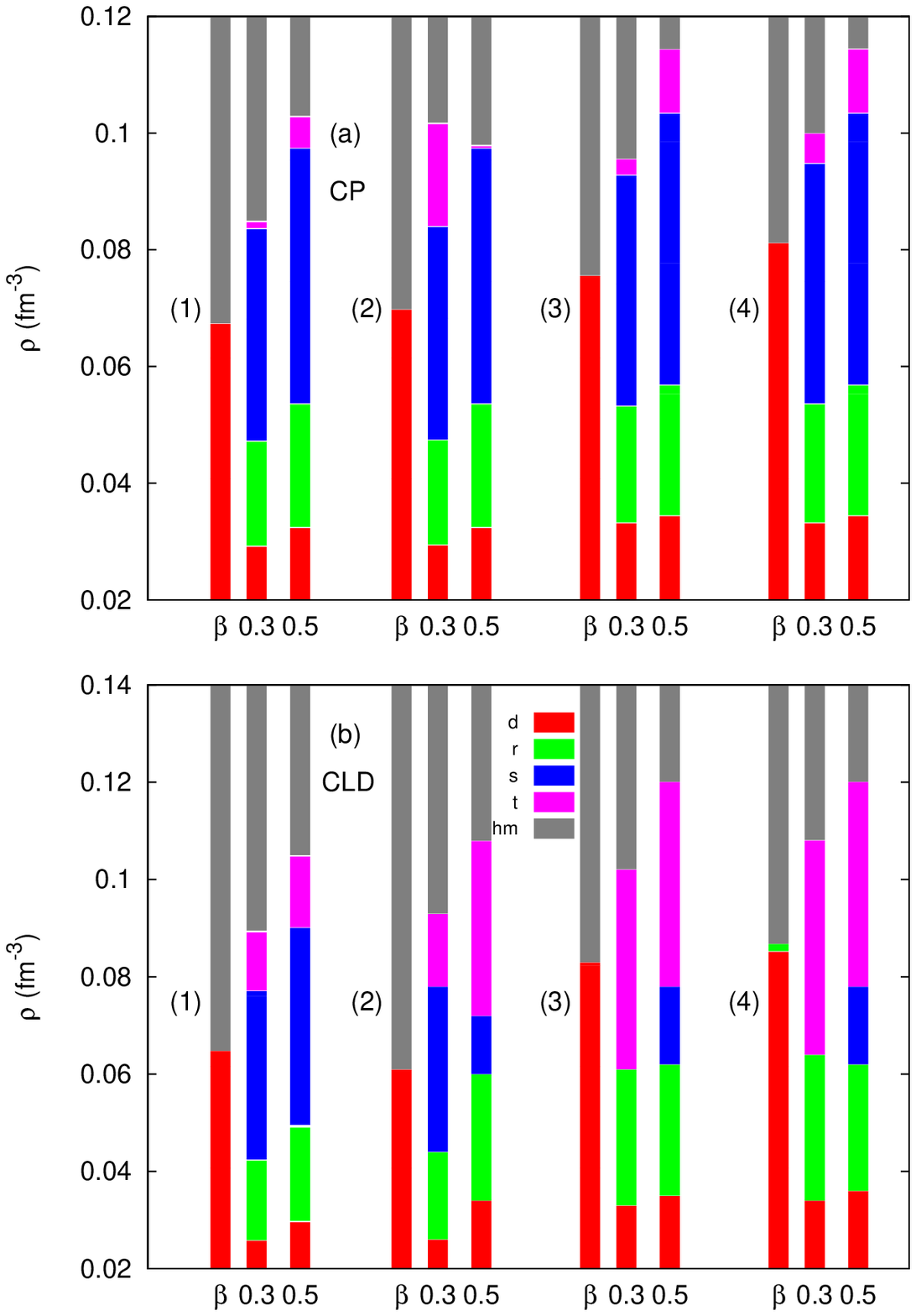}
	\caption{(Color online) Pasta phases for the eNJL1 (1), eNJL1$\omega\rho$1 (2), eNJL3 (3), and eNJL3$\sigma\rho$1 (4) interactions within the CP (a) and CLD (b) calculations for $\beta-$equilibrium, $y_p=0.3$ and $y_p=0.5$ matter. The temperature is fixed to 0 MeV.}
\label{fig7}
\end{figure}

We next compare the behavior of neutron matter, calculated with
  the models under study, with microscopic calculations.
 In Fig. \ref{fig5}, we show the pressure as a function of the
  density for pure neutron matter obtained with the eNJLx,
  eNJLx$\omega\rho$y  and the eNJLxm$\sigma\rho$  models. 
The two colored bands in the figure are the
  results from microscopic calculations based on chiral NN and 3N
  interactions (green) \cite{Hebeler-13} and Quantum Monte Carlo
  results (gold) \cite{Gandolfi12}. In general, almost all  models
  are stiffer than predicted by the microscopic calculations. Only
  eNJL1$\omega\rho$2, eNJLxm$\sigma\rho$1, and eNJL3$\sigma\rho1$
  models are inside  the constrained region for a wide range of 
densities. 

 In the following,  we will not consider anymore the
  eNJL1$\omega\rho$2 model because, although it presents good properties at
  saturation and subsaturation densities, at
  suprasaturation  it fails, not  only to
  satisfy  heavy-ion flow constraints, but even worse, it predicts a
  transition to pure neutron matter at a too low density and
  therefore, is not appropriate to describe stellar matter.  

\subsection{Subsaturation EoS}

At subsaturation densities, homogeneous nuclear matter is not stable
and matter has the tendency to clusterize. Applying the formalisms, CP
and CLD, described in section \ref{CP-CLD}, we present in Fig. \ref{fig6} the
energy per particle and the pressure versus density obtained  for
homogeneous and clusterized matter, and three different types of
matter, $\beta$-equilibrium matter and matter with a proton fraction
equal to 0.5 and 0.3. 
As expected, clusterized matter has a smaller energy per particle. The CLD and CP
methods give similar results, CP giving rise to a slightly smaller
energy per particle due to the fact that the Coulomb field
and surface energy contributions are included after the minimization
of the free energy. For $y_p=0.3$ and 0.5, there is a first order phase
transition between the clusterized phase and the homogeneous phase.

The different geometries that are present in the non-homogeneous phase
depend on the proton fraction, model and method. In Fig. \ref{fig7},
the distribution of the different types of shapes as a function of the
density are given for four models, eNJL1, eNJL1${\omega\rho}1$, eNJL3
and eNJL3$\sigma\rho$1, 
and the two methods, CP and CLD. In all models, $\beta$-equilibrium
matter does not contain exotic shapes, and clusters are all spherical,
independently of model and method,  except for a very small region
  where rods appear for eNJL3$\sigma\rho1$. 
This behavior has   been obtained
in \cite{Oyamatsu-07}, for models with a large symmetry energy slope at
saturation density. In their analysis, $L\lesssim 80$ MeV was the condition
to obtain other shapes besides droplets.

Matter with $y_p=0.3$ and 0.5 have, besides droplets, rods, slabs and
tubes. Bubbles are never present. Generally, the CLD predicts an earlier
transition to homogeneous matter and larger tube regions, but this is not always the case.

\begin{figure}[!htbp]
	\begin{tabular}{c}
	   \includegraphics[width=0.5\textwidth]{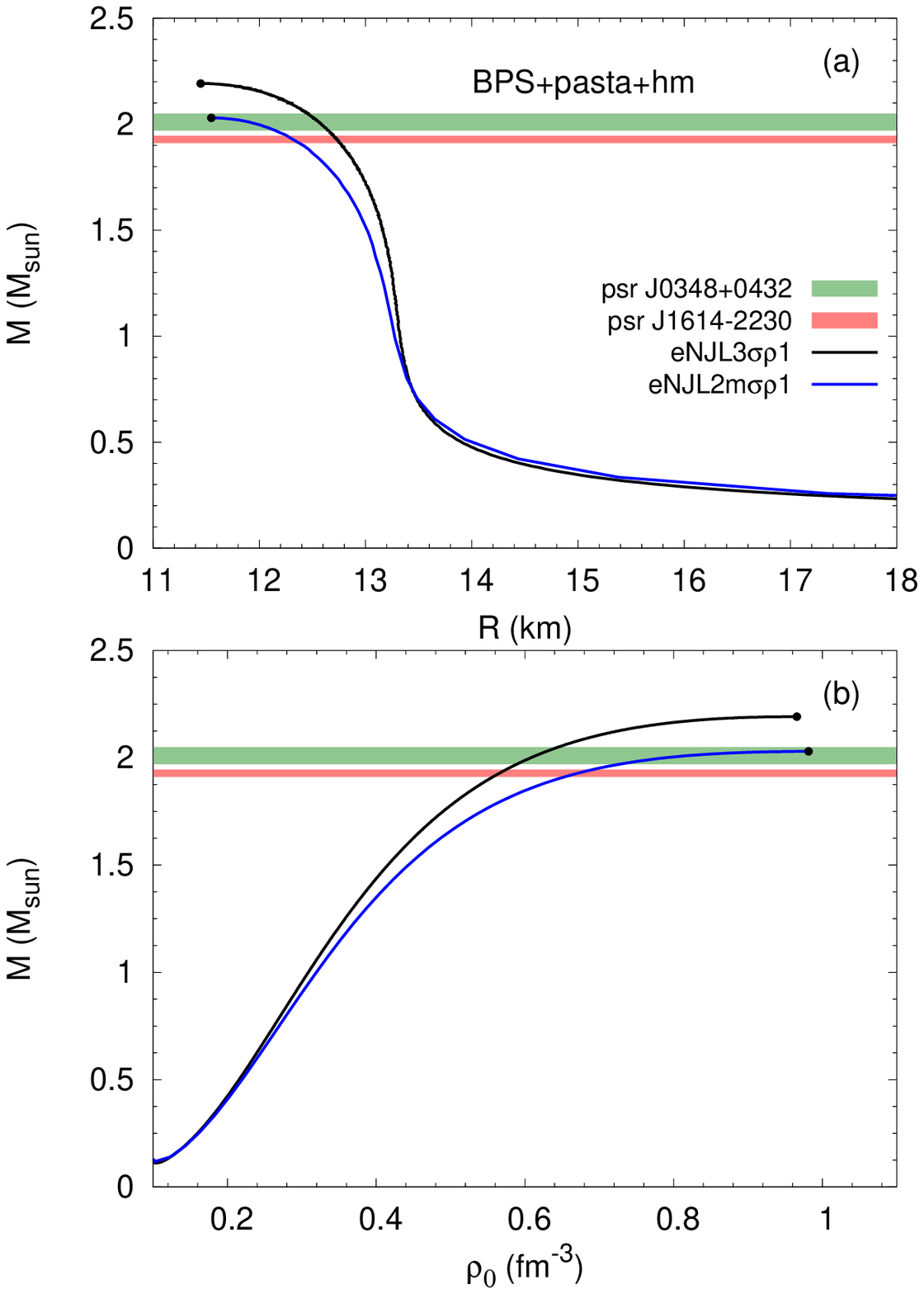}
	\end{tabular}   
	\caption{(Color online) Mass-radius relation (a) and mass as a function of the central density (b) for the family of hadronic stars that passed the imposed constraints. We have considered pasta in the EoS. The outer crust is  given by the BPS EoS. The black dots correspond to the  maximum mass star.}
\label{fig8}
\end{figure}

\begin{table*}[!htbp]
  \centering
  \caption{Some properties of the families of hadronic stars considered. The masses, $M$, the radii, $R$, and the densities, $\rho$,  are given in units of $M_{\odot}$, km, and fm$^{-3}$, respectively. * These models contain broken chiral symmetry up to the densities considered.} 
  \begin{tabular}{c c c c c c c c}
    \hline
    \hline
	Model & $M_{g_{max}}$  & $M_{b_{max}}$ & $R$  & $\rho_c$   & $\rho_\chi$  & $R_{M=1.4M_\odot}$  & $\rho_{c_{M=1.4M_\odot}}$  \\
    \hline
	  		  	with pasta  & & &&&&&\\
	eNJL1 (CP) &	2.607  &3.177  &12.397  &0.776 &  0.35 & 14.406 & 0.324 \\
	eNJL1 (CLD) &	2.607  &3.177  &12.386  &0.776 &  0.35 & 14.406 & 0.324 \\
	eNJL1$\omega\rho1$  &	2.520  & 3.052  &12.427  &0.796 & 0.35 & 14.198 & 0.343 \\
	eNJL2  &	 2.365 & 2.835 & 11.556 & 0.916 &  0.45 & 14.086 & 0.350  \\
	eNJL2$\omega\rho1$ & 2.195 & 2.592 & 11.926 & 0.891  & 0.45 & 13.786 & 0.383  \\
	eNJL3  &	 2.289 & 2.714 & 12.057 & 0.875 &  0.998 & 13.931 & 0.339  \\
	eNJL3$\sigma\rho1$ & 2.192 & 2.602 & 11.445 & 0.966  & 0.998 & 13.212 & 0.391 \\
	eNJL1m  &	 2.072 & 2.409 & 12.398 & 0.866 & * & 13.839 &  0.356 \\
	eNJL1m$\sigma\rho1$ & 1.884 & 2.180 & 11.521 & 1.013  & * & 12.809 & 0.449  \\
	eNJL2m  &	 2.275 & 2.683 & 12.380 & 0.846 &  * & 14.168 & 0.326  \\
	eNJL2m$\sigma\rho1$ & 2.030 & 2.375 & 11.549 & 0.981  & * & 13.084  & 0.414 \\
    \hline
	  		  	without pasta  & & &&&&&\\
	eNJL1  & 2.607 & 3.177 & 12.297 & 0.776 &  0.35 & 13.971 & 0.324 \\
	eNJL1$\omega\rho1$  & 2.520 & 3.052 & 12.326 & 0.796 & 0.35 & 13.773 & 0.343 \\
	eNJL2  & 2.365 & 2.835 & 11.477 & 0.916 & 0.45 & 13.708 & 0.350 \\
	eNJL2$\omega\rho1$  & 2.195 & 2.592 & 11.826 & 0.891  & 0.45 & 13.456 & 0.383 \\
	\hline
    \hline
  \end{tabular}
 \label{tab3}
\end{table*}

\subsection{Neutron stars}
\label{ns}

We have built an EoS, appropriate to describe cold stellar matter
in $\beta$-equilibrium, as explained in the following: a) for the outer crust, the
Baym-Pethick-Sutherland (BPS) \cite{bps} EoS was taken; b) the inner crust EoS
was obtained within the CP method, and, for reference, we will
also show results where the homogeneous matter EoS was taken for
densities above neutron drip c) the core of nucleonic stars is described
using  all models, except eNJL1$\omega\rho2$, 
none of which predicts the transition to neutron matter at a
density below the central density of the most massive star; d) for the
core of hybrid stars, eNJL1 is considered for the hadronic phase,
and the quark phase is described within the su(3) NJL model.

\begin{figure}[!htbp]
   \includegraphics[width=0.45\textwidth]{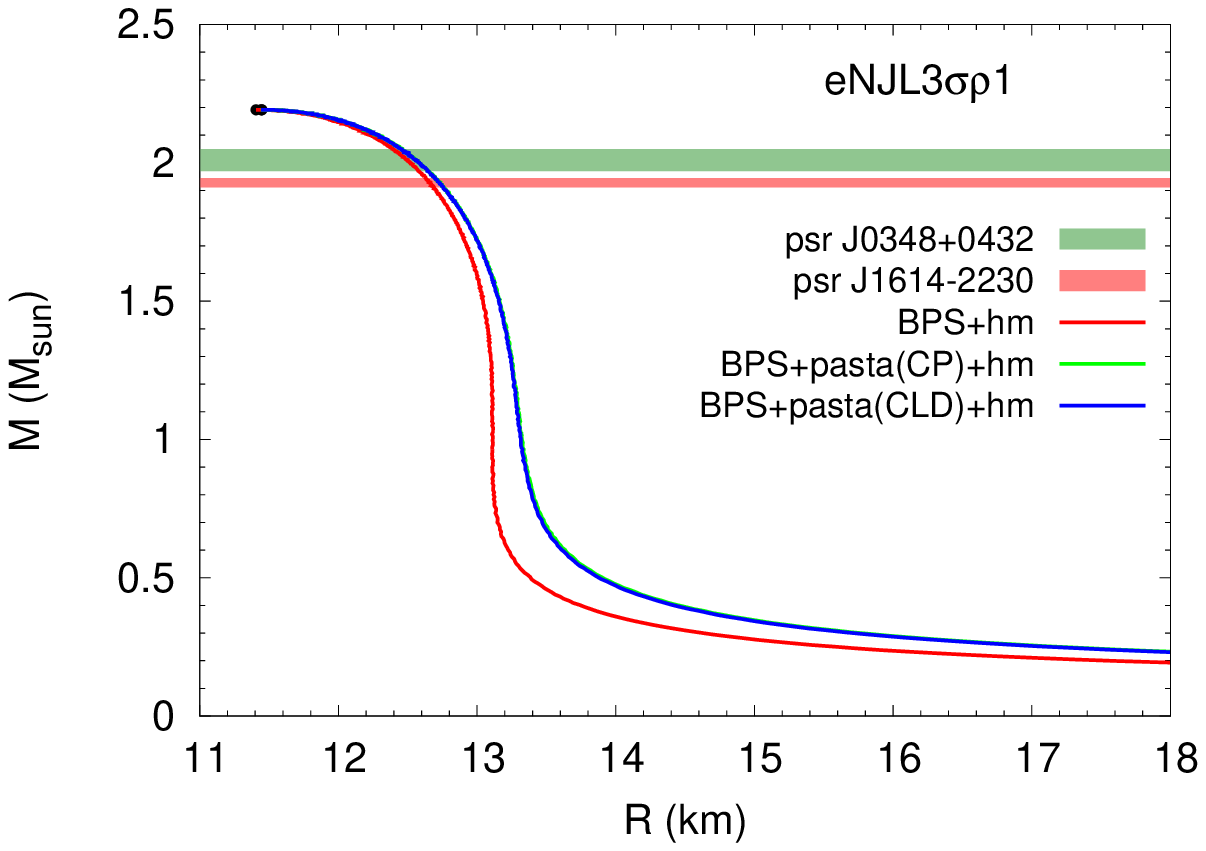}
	\caption{(Color online) Mass-radius relation for the eNJL3$\sigma\rho$1
          model, with pasta from a CP (green) and a CLD (blue) calculations in the EoS, and without
          it (red).  The outer crust is
          given by the BPS EoS. The black dots correspond to the
          maximum mass star.}
\label{fig9}
\end{figure}

The $M(R)$ curves of the nucleonic stars are plotted in Figures \ref{fig8} and \ref{fig9},
and some of their properties are given in Table \ref{tab3}.  These
properties are the maximum gravitational and baryonic masses and
corresponding radii, the central and chiral symmetry restoration
densities, the radius of a 1.4$M_\odot$ star, and its respective
central density. It is seen that  the maximum mass and central
densities do not depend on how the inner crusts are described, but the
radius is sensitive to the crust EoS, a larger radius being obtained
when the pasta is included, see Fig.  \ref{fig9}. This
  difference can be as high as 500m for a 1.4 $M_\odot$ star, and 1km
  for a 1.0 $M_\odot$ star.  The $\beta-$equilibrium
  EoS of the inner crust, obtained both within the CLD and CP
  approaches, are very similar, giving rise to  almost identical radii, as seen in Fig. \ref{fig9}. 

Maximum masses are all above two solar masses, satisfying the constraint imposed by pulsars  PSR J0348+0432 \cite{Antoniadis13} and PSR J1614-2230 \cite{Demorest10,Fonseca16},. All models predict a central core of matter for the maximum mass star with the chiral symmetry restored, except for eNJL3 and eNJL3$\sigma\rho$1. However, the $1.4 M_\odot$ stars are formed only of matter in the broken phase. Confirming previous results of
\cite{Carriere03}, 1.4 $M_\odot$ stars have a smaller radius for
smaller slopes $L$.  Their radii are quite large, and the smallest
one, obtained including the pasta in the inner crust, is 13.21  km, for a model without current mass in the EoS. 
The two models with the same value of $L$ have different incompressibilities, and the larger radius corresponds to the
larger $K$. Stars within eNJL1 models have larger maximum masses
because they have stiffer EoS, see Fig. \ref{fig2}. In the right
panels of Fig.\ref{fig8}, we show the $M/R$ relation and mass as a
function of the density for the models with current mass in the
EoS. eNJL1m$\sigma\rho1$ is the only one that does not produce a maximum mass star of $2 M_\odot$. All the other models are within the observational constrains.

 In Table \ref{tab5}, we show which models satisfy / do not satisfy the
 experimental and observational constraints and the microscopic calculations (\cite{Hebeler-13, Gandolfi12}).

\begin{table*}[!htbp]
  \centering
  \caption{ The models considered and the experimental, observational, and microscopic calculations constrains.}
  \begin{tabular}{c c c c c c}
    \hline
    \hline
	Model & \multicolumn{2}{c} {Experimental}  & \multicolumn{2}{c}{Microscopic} & Observational  \\
	\cline{2-6}
	& Kaons & Flow & $\chi$-NN,3N & QMC & 2 M$_\odot$ \\
    \hline
	eNJL1 &yes &no &no &no &yes \\
    eNJL1$\omega\rho$1 & yes & no & no & no &yes \\
    eNJL1$\omega\rho$2 & yes & no & yes & yes & no \\
    	\hline
    eNJL2 & yes & practically & no & no & yes  \\
    eNJL2$\omega\rho$1 & yes & practically & no & no & yes \\
    \hline
    eNJL3 & yes & yes & no & no & yes \\
    {\bf eNJL3$\sigma\rho$1} & {\bf yes} & {\bf yes} &  {\bf yes} & {\bf yes} & {\bf yes} \\
    \hline
    eNJL1m & yes & yes & no & no & yes \\
    eNJL1m$\sigma\rho$1 & yes & yes & yes & yes & no \\
    \hline
    eNJL2m & yes & yes & no & no & yes \\
    {\bf eNJL2m$\sigma\rho$1} & {\bf yes} & {\bf yes} & {\bf yes} & {\bf yes} & {\bf yes} \\
	\hline   
    \hline
  \end{tabular}
 \label{tab5}
\end{table*}

\begin{figure*}[!htbp]
   \includegraphics[width=\textwidth]{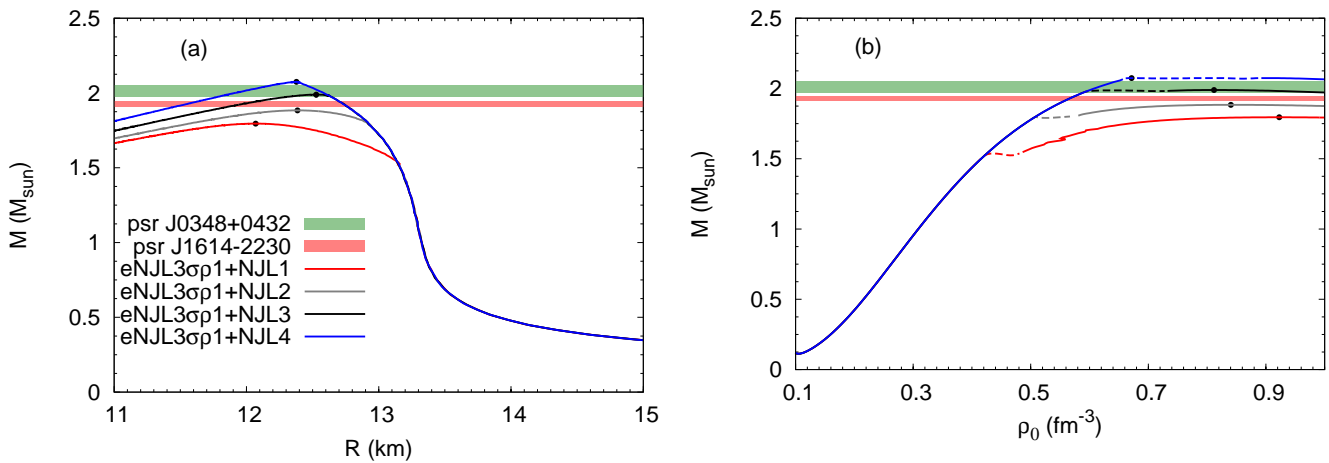}
	\caption{(Color online) Mass-radius relation (a) and mass as a function of the central density (b) for the hybrid stars families. Pasta is included in the hadronic EoS. The black dots correspond to the maximum mass star, and the dashed lines in the left panel correspond to the mixed hadron-quark phase in the stars.}
\label{fig10}
\end{figure*}

We have next built a hybrid star EoS, using the Maxwell construction to
match the hadronic and quark EoS. Quark matter is described within a new parametrization for the
su(3) NJL  model \cite{Pereira16}, with a vector interaction, and  several strengths of this
interaction are considered.  Results are presented in Table
\ref{tab4}  and Fig. \ref{fig10}. 
 We observe that  a core of quark matter is obtained  for all the
 cases considered, except for the largest vector coupling (NJL4), where the onset of the quark phase makes the star unstable. This means that
   stable stars are purely hadronic, or at most, contain in their core a
   mixed quark-hadron phase. The other three  models result in maximum
   masses below 2$M_\odot$, but the one with $G_v=0.12 G_s$  is above the measured mass  1.928 $\pm 0.017 M_\odot$ for PSR J1614-2230 \cite{Fonseca16} and within  the limits of the  2.01$\pm$0.04 $M_\odot$ mass of the PSR J0348+0432 \cite{Antoniadis13}. There are other cases where NJL-type models were used for the quark matter EoS, and where massive hybrid stars were found, see e.g. Ref. \cite{Benic15} and references therein.

\section{Conclusions} \label{IV}

In the present work, we have described nuclear and stellar matter
within a relativistic nuclear model with chiral symmetry. At  first, three
parametrizations with different isoscalar properties were considered, 
having different onset densities for the restoration of
chiral symmetry. These models present a quite large symmetry energy
slope at normal density and, therefore, four other models have been
proposed with a smaller slope.  To accomplish this new feature, a
mixed vector-isovector - vector-isoscalar term, or a mixed scalar-isoscalar - vector-isovector term, were included in the Lagrangian density.
 Above the restoration of chiral symmetry,
the EoS of symmetric nuclear matter for  all models becomes much
stiffer. On the other hand, the symmetry energy softens above the chiral
symmetry  restoration  density and, at large enough densities, it may even
become negative, mainly if the modification of the density
  dependence of the symmetry energy is accomplished by including
  isovector vector-isoscalar vector mixed terms.  A special case is
  the eNJL3, in which the
restoration of the chiral symmetry 
happens at a very high density, $\rho\sim 7\rho_0$. 
We have also  implemented two other models,
where we considered a current mass in order to soften the restoration
of the chiral symmetry,  making the EoS less stiff.  A
mixed scalar-isoscalar - vector-isovector term was added to the models in order
to decrease the slope of the symmetry energy even further, allowing a 
softening in the EoS.

Neutron star radii are still not well constrained, and  it is expected that the future X-ray telescopes, like NICER and Athena, will impose much stronger constraints. Our results are compatible with some of the present predictions.
 Both hadronic and hybrid star radii of 1.4 $M_\odot$ are above 12.8 km, within the observations of
the objects BNS 4U 1608-522 \cite{Poutanen14}, BNS SAX J1748.9-2021 \cite{Guver13, Heinke14}, and RP-MSP PSR J0437-4715 \cite{Bogdanov13, Verbiest08}, but out of the range $10.1-11.1$ km obtained in \cite{Ozel16}, from the analysis of spectroscopic radius measurements of twelve neutron stars obtained during thermonuclear bursts or in quiescence. However, in \cite{Chen15}, it was shown that in order to prevent the EOS from violating causality, the radius should satisfy $R_{1.4}\gtrsim 10.7$ km, if it is imposed that the EOS  also describes a 2 $M_\odot$ star. In \cite{Steiner16}, taking experimental constraints and causality restrictions for large maximum masses,  the $1.4 M_\odot$ star radii  were constrained to be within the interval 12.1$\pm$ 1.1 km,   (see \cite{Fortin15}).

\begin{table*}[!htbp]
  \centering
  \caption{Some properties of the families of hybrid stars considered. The masses, $M$, the radii, $R$, and the densities, $\rho$,  are given in units of $M_{\odot}$, km, and fm$^{-3}$, respectively. The vector interaction coupling constant, $G_v$, for the NJL model is also shown.}
  \begin{tabular}{c c c c c c c c}
    \hline
    \hline
	Model & $G_v/G_s$ (NJL) &$M_{g_{max}}$  & $M_{b_{max}}$ & $R$  & $\rho_c$ & $R_{M=1.4M_\odot}$  & $\rho_{c_{M=1.4M_\odot}}$  \\
    \hline
	eNJL3$\sigma\rho$1+NJL1 &0 &1.795 &	2.051 & 12.069 & 0.923 & 13.211 & 0.392 \\
	eNJL3$\sigma\rho$1+NJL2 & 0.05&1.884 & 2.166 & 12.386 & 0.840 & 13.211 & 0.392 \\
	eNJL3$\sigma\rho$1+NJL3 & 0.12 & 1.989 & 2.311 & 12.527 & 0.812 & 13.211 & 0.392  \\
	eNJL3$\sigma\rho$1+NJL4 & 0.2 & 2.074 & 2.430 & 12.378 & 0.671 & 13.211 & 0.392 \\
	\hline   
    \hline
  \end{tabular}
 \label{tab4}
\end{table*}

 The EoS of non-homogeneous subsaturation matter was built
within the CP and the CLD methods, and we found that within these models,
$\beta$-equilibrium matter does not present a non-spherical pasta phase, except for eNJL3$\sigma\rho1$.
Non-spherical shapes will, however, occur for larger proton fractions,
and could exist in core-collapse supernova matter. However, to
describe this kind of matter, a finite temperature calculation must be
performed.

Having the subsaturation EoS of $\beta$-equilibrium matter, we have
built an almost unified hadronic stellar matter EoS, with the outer crust
described by the BPS EoS, and the inner crust and core described
within the eNJL model. It was shown that an uncertainty of 0.5
  and 1 km in the radius, respectively, of a 1.4 and a 1.0 $M_\odot$
  star is obtained when the homogeneous matter EoS is used to describe
  the inner crust. 

For the core, we have considered not only
nucleonic matter, but also a possible  phase transition to
quark matter, described within the su(3) NJL model. In the quark model,
we have included a vector term that allows to turn the quark EoS
stiffer.
All nucleonic star families obtained with the models that do not
predict a neutron instability for densities below the central density,
the maximum mass obtained is well above 2$M_\odot$. 
The inclusion of a possible deconfinement phase transition 
either decreases the maximum mass (as expected) to values below 2
$M_\odot$,  but still within the mass constraints imposed by the pulsars  PSR J1614-2230 \cite{Demorest10,Fonseca16} and PSR J0348+0432 \cite{Antoniadis13} mass,  or renders the star with a quark core unstable.

\section*{ACKNOWLEDGMENTS}
H.P. is supported by FCT under Project No. SFRH/BPD/95566/2013. She is very thankful to D.P.M. and the Departamento de F\'isica of the Universidade Federal de Santa Catarina for the kind hospitality during her stay there. She is also grateful to M.B. Pinto and F. Gulminelli for useful discussions. Fruitful discussions with Pedro Costa are acknowledged. D.P.M. acknowledges support by CNPq (grants 300602/2009-0) and FAPESC (Brazil) under project 2716/2012, TR 2012000344. Partial support comes from ``NewCompStar'', COST Action MP1304.

\end{document}